# Rule-based Adaptations to Control Cybersickness in Social Virtual Reality Learning Environments

Samaikya Valluripally, Vaibhav Akashe, Michael Fisher*, David Falana+, Khaza Anuarul Hoque, Prasad Calyam
University of Missouri-Columbia, *Columbia College, +Rutgers University
{svbqb, vva8yb}@mail.missouri.edu, {mefisher2}@cougars.ccis.edu, {falanadavid}@gmail.com, {hoquek, calyamp}@missouri.edu

*Abstract*—Social virtual reality learning environments (VR-LEs) provide immersive experience to users with increased accessibility to remote learning. Lack of maintaining high-performance and secured data delivery in critical VRLE application domains (e.g., military training, manufacturing) can disrupt application functionality and induce cybersickness. In this paper, we present a novel rule-based 3QS-adaptation framework that performs risk and cost aware trade-off analysis to control cybersickness due to performance/security anomaly events during a VRLE session. Our framework implementation in a social VRLE viz., vSocial monitors performance/security anomaly events in network/session data. In the event of an anomaly, the framework features rule-based adaptations that are triggered by using various decision metrics. Based on our experimental results, we demonstrate the effectiveness of our rule-based 3QS-adaptation framework in reducing cybersickness levels, while maintaining application functionality. Using our key findings, we enlist suitable practices for addressing performance and security issues towards a more high-performing and robust social VRLE.

*Index Terms*—Social Virtual Reality, IoT/Cloud Middleware, Cybersickness, Queuing Model, Control Adaptation

## I. INTRODUCTION

Social Virtual Reality Learning Environments (VRLEs) are a convergence of virtual reality (VR), Internet-of-Things (IoT) and cloud computing technologies [1]. As shown in Figure 1, they integrate real-world smart things (i.e., VR headsets/glasses) with virtual objects/avatars for a real-time immersive interaction of geographically distributed users [2]. Social VR applications in education or collaborative tasks adopt virtual worlds as learning environments [3], [4], where participants can interact effectively with higher engagement and performance [5]. To facilitate continuous interaction between the users (e.g., instructors and students), the networked VRLE components collect data from distributed user locations, and seamlessly integrate web-based tools to render VRLE content. However, such capabilities in these socio-technical systems demand for high-performance and robust VRLE application features.

With the dynamic user-system interactions for content rendering, VRLEs are a target for an attacker to trigger security attacks [6], [7]. In addition, the work in [8] details about the performance issues that can disrupt the social VRLE user

This material is based upon work supported by the National Science Foundation under Award Numbers CNS-1647213, CNS-2114035, and CNS-1950873. Any opinions, findings, and conclusions or recommendations expressed in this publication are those of the authors and do not necessarily reflect the views of the National Science Foundation.

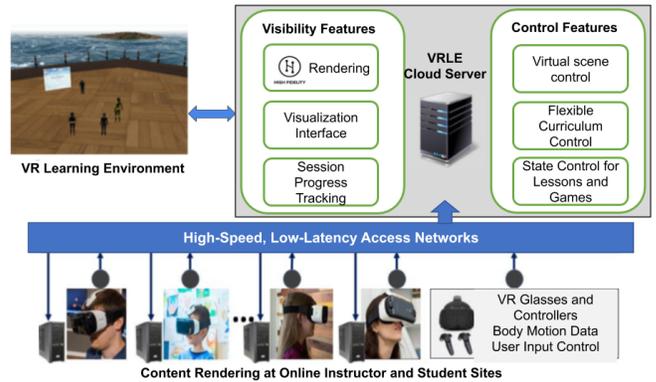

Fig. 1: Illustration of a cloud-based Social VRLE deployed across high-speed, low-latency network sites.

experience. However, prior works lack in the knowledge to address both performance and security issues that can impact the user experience and user safety in VRLE sessions. Failure to address such impediments can lead to deface attacks on the VR content with offensive images [9] that can hamper user experience. They can also lead to application latency issues that degrade performance. Based on prior works in VRLE and other IoT applications [10], [11] we adopt the following definitions of various performance ($3Q$) factors: Quality of Application (QoA) – a measure of the application performance; Quality of Service (QoS) – a measure of network resources such as bandwidth and jitter; Quality of Experience (QoE) – a measure of the perceived satisfaction or annoyance of a user's experience [12]. Similarly, we adopt the definition of security – as a condition that ensures a VR system is able to perform critical application functions with the establishment of confidentiality, integrity, and availability [7]. Together, such performance and security issues can induce "cybersickness" (e.g., eyestrain, nausea, headache, disorientation of user movement) [13], [14]. Hence, there is a need to study methods to mitigate impact of performance and security anomaly events that induce cybersickness.

In this paper, we propose a novel rule-based 3QS-adaptation framework that controls cybersickness induced by both performance i.e., {QoS, QoA, QoE} (3Q) factors and security i.e., {Denial of Service (DoS), intrusion} (S) issues which are further termed as 3QS issues of a social VRLE application. Our framework design and implementation are motivated by an exemplar social VRLE i.e., vSocial [1] that provides access to online VR-based content to enhance learning effectiveness of geographically dispersed VRLE users. Through continuous monitoring of performance and security anomaly events in VRLE sessions with a web-based user interface that we developed, we derive a priority-based queuing model that guides adaptive control of cybersickness levels. In this context, we objectively measure cybersickness [15] in terms of a *(latency metric)* using experimental validation and simulation results. Next, we determine the suitable adaptation for a given anomaly event, using a novel dynamic decision making mechanism using metrics such as: *(impact on cybersickness level, history of adaptations, response time)*. The choice of adaptation is made by performing a risk level calculation along with the cost associated with the adaptation in order to reduce the cybersickness levels due to 3QS issues.

We demonstrate the effectiveness of our 3QS-adaptation framework in vSocial using an IoT-cloud based testbed setup that involves vSocial on Amazon Web Services (AWS) resources. Our validation results show that our adaptation choices are effective in reducing the cybersickness levels and in maintaining the application functionality at a usable level. Our experimental results are further used to create a knowledge base for a social VRLE application that includes user data, session information, anomaly event data and their respective implemented adaptations along with their associated usability metrics. The knowledge base can also be used for training the decision making process to handle future anomaly events in Social VRLE sessions. Lastly, using our key findings, we enlist the suitable practices for addressing performance and security issues towards a more secure and operational social VRLE following the NIST SP800-160 [16], [17] principles and AWS recommendations [8], [18].

The remainder of the paper is organized as follows: Section II discusses related work. Section III presents the proposed rule-based 3QS-adaptation framework. Section IV details the priority-based queuing model used in the adaptation framework. Section V presents the performance evaluation in a IoT/cloud testbed featuring vSocial. Section VI concludes the paper and suggests future work.

## II. RELATED WORK

### A. Factors Impacting VRLE Applications

Prior works [6]–[8] addressed performance and security issues in social VRLE applications. The work in [6] described potential security, privacy and safety issues that can trigger disruption in the VRLE application functionality. In addition, the work in [7] also detailed vulnerable components in VRLE that can lead to sophisticated cyber-attacks such as Loss of Integrity and privacy leakage. Authors in [8] model performance issues via a 3Q-model to determine the causes of disruption of VRLE user experience.

The impact of such effects can specifically induce cybersickness [19]–[21], thus compromising *user safety* in a VRLE session [13], [14]. On the other hand, works related to other applications such as remote instrumentation [22] and video-based cloud applications [23] analyze performance factors that disrupt user experience and propose a 3Q factors interplay model for determining suitable adaptations. Using the outlined security and performance issues of VRLE in the above state-of-the-art, we propose a continuous 3QS anomaly event monitoring approach to guide adaptation control decisions to minimize cybersickness levels during a VRLE session.

### B. Application Adaptation Frameworks

Prior works [20], [24], [25] proposed approaches that focus on detection of cybersickness for users in a VRLE. The adaptation mechanisms discussed in [24], [25] focused on improving the VRLE application features. There have been works [26], [27] that address either performance or security issues in the context of a control-feedback scheme to adapt cloud-based IoT applications. For instance, the works in [26]–[28] present solutions that feature adaptive control mechanisms to address scalability and latency issues based on user's service level objective (SLO) and cost constraints. Adaptive control mechanisms [29] related to addressing security issues at the application layer have been studied at an on-demand resource management level involving e.g., DoS attacks [30]. In contrast, our 3QS-adaptation framework considers the interplay of security and performance factors potentially inducing cybersickness. Our adaptations consider time-sensitive response of the system by using performance metrics such as: response time, resource usage for an adaptation and risk of performing that adaptation along with the cost constraint for a given performance/security issue.

Our work builds upon prior works such as [31], [32] that derive analytical models to aid in resource allocation issues for cloud-based IoT applications. The work in [31] uses a *Multi-stage Queuing Model* with an embedded Markov chain process to evaluate application QoS in order to improve QoE of users. Whereas, the authors in [32] develop an analytical model which uses the Markov chains and M/M/1/K queuing system to provide close-form expressions to characterize elasticity of cloud-based firewalls. Our work uses similar assumptions of cloud-based IoT applications, and we choose a priority-based queuing model to process anomaly events that are inducing higher cybersickness levels. Our choice of priority-based queuing model is to avoid waiting delays of processing the utmost severe anomaly events that could lead to a higher chance of increase in cybersickness for users.

## III. RULE-BASED 3QS-ADAPTATION FRAMEWORK

In this section, we present the overview of our novel rule-based 3QS-adaptation framework to control the impact of cybersickness levels in a VRLE as shown in Figure 2. Firstly, we detail the monitoring process for detecting 3QS anomaly events in the collected network data during a VRLE session. Next, we describe how our decision module determines suitable adaptations relevant to these anomaly events. Following this, we explain how the decision outcome is incorporated on an affected VRLE component such that the cybersickness level is reduced. Lastly, we describe how we update session

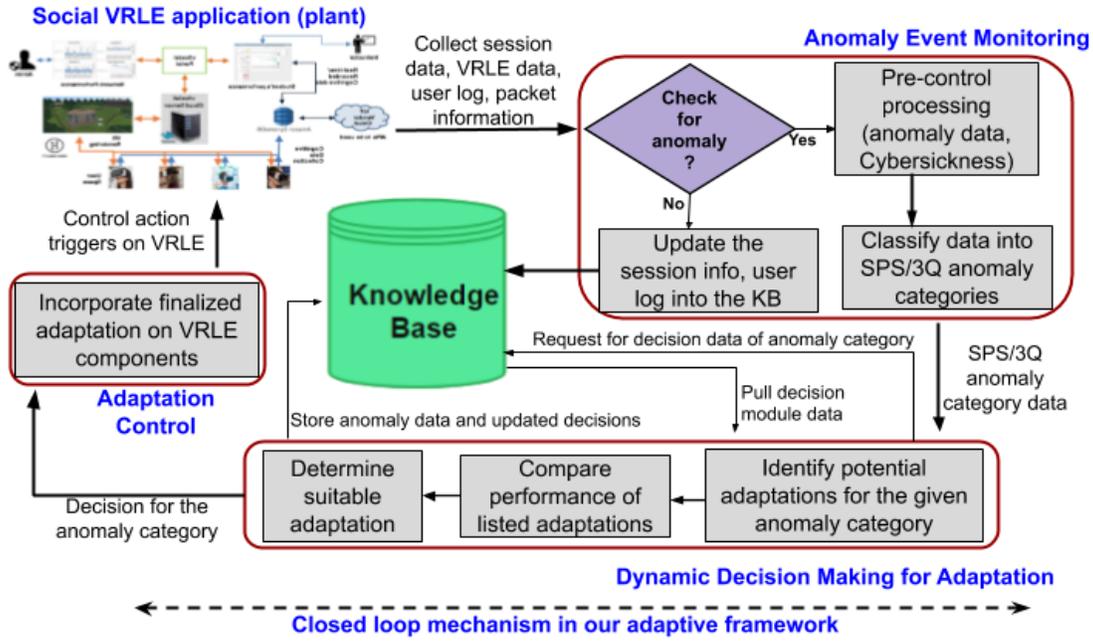

Fig. 2: Proposed rule-based 3QS-adaption framework for a social VRLE system.

information and the impact of the adaptation into a knowledge base.

### A. Anomaly Event Monitoring Tool

To identify any potential 3QS issues in a social VRLE, we developed an anomaly event monitoring tool [33] building upon the work in [34] to observe the network behavior changes, and user activity trends during the VRLE session. We create alarms to trigger when an anomalous behavior pattern is identified in the vSocial application. The anomaly event types include: QoA issues (e.g., visualization delay due to network lag), QoS issues (e.g., packet loss), and security issues (e.g., DoS attack, unauthorized access). Next, we collect this anomaly event data as shown in Figure 2 in order to calculate the corresponding impact on the cybersickness level for the session user(s). Following this, we classify the collected anomaly event data into specific 3QS categories.

**Algorithm 1** Build Decision Units

**Input:** Anomaly type list
**Output:** Relevant Decision units
1 **begin**
2     **Function** $BuildAdaptation$ ():
3         **for** each $AnomalyType \in AnomalyTypeList$ **do**
4             **Function** $BuildDecisionUnits$ ():
5                 Let $TupleList = [\,]$
6                 **for** each $Adaptation \in TupleList$ **do**
7                       Let $Tuple = (A_n, Ct, I)$ $TupleList.append(Tuple)$
8                 **end**
9                 **return** $DecisionUnit\{AnomalyType_i, TupleList\}$
10             **end Function**
11         **end**
12         **return** $Adaptation\{DecisionUnit_0, ..., DecisionUnits_k\}$
13     **end Function**
14 **end**

### B. Adaptation Decision Making

The anomaly event data is classified based on 3QS issue categories. Given anomaly event, our 3QS-adaptation framework activates a decision module that has knowledge of potential adaptations for the specific event as shown in Figure 2. Each of the detected anomaly event categories are sent as input to the Algorithm 1 which details the functionality of the decision module. The decision module allows it to compare an anomaly event in a particular category with a set of relevant decision units as described in Algorithm1. Each decision unit has the knowledge on how to deal with a specific type of anomaly event i.e., decision units contain a list of potential candidate adaptations that are retrieved from the knowledge base module. The function *BuildDecisionUnits()* in Algorithm 1 describes how decision units are developed where, each decision unit contains a list of defined tuples. These tuples are of the form $\{A_n, Ct, I\}$, where $A_n$ represents the adaptation name, $Ct$ represents the history of adaptation in terms of number of times that specific choice was implemented, and $I$ represents the impact on cybersickness level after the adaptation was implemented for a given anomaly event. As and when such decision units are created, our decision module retrieves the decision units using the *BuildAdaptation()* function in Algorithm 1.

Next, the decision module traverses through the list of candidate adaptations in each of these retrieved decision units to determine the most suitable adaptation. Each of the listed decision units will be sorted using the order of attributes $I$, $Ct$ in tuples which are termed as "decision metrics" along with the reduced response time taken by a specific adaptation. The head of the sorted list of candidate adaptations represent the most suitable adaptation for a given anomaly event. With every iteration of handling anomaly events, the $Ct$ value related to the considered adaptation gets updated into the knowledge base. In addition, our algorithm avoids the conflict scenarios for different adaptations given by the decision units. For this, our framework considers the anomaly events as part of a priority queue. We further consider each anomaly issue as an independent event to avoid any conflicts of adaptations. Thus, using the decision module, our 3QS-adaptation framework facilitates dynamic decision making for a suitable adaptation to reduce the induced cybersickness level for a given anomaly event.

*C. Creation of the Knowledge Base*

Our proposed 3QS-adaptation framework stores the baseline data of benign application behavior into the knowledge base for handling future anomaly events in a social VRLE. The knowledge base actively stores VRLE session information, detected anomaly event patterns along with the potential adaptations and associated user data. The anomaly event traces in the knowledge base can be helpful to a network/system administrator to determine the causes of the detected anomalies, and improve the effectiveness of the adaptations. Moreover, the knowledge base can be used as a medium for threat intelligence collection to train our decision module for mitigation of zero-day 3QS anomaly events that can arise in an individual scenario and/or in combination scenarios.

*D. Adaptation Control*

The control module in our 3QS-adaptation framework enacts suitable adaptations for a given anomaly event category. Once the decision outcome (i.e., suitable adaptation) is obtained, the control module first calculates the risk level associated to a choice of adaptation, along with the cost incurred to control the induced cybersickness anomaly event. Next, the control module invokes an action using an alarm (using e.g., AWS CloudWatch) for the relevant functionality of the determined adaptation. In addition, the risk and cost aware decision outcome implementation is evaluated for the feedback (e.g., control on cybersickness level, user satisfaction). If the anomaly event is successfully handled, then this session information along with the control module data is updated into the knowledge base for handling similar future anomaly events. Thus, the anomalies are monitored continuously, and we perform dynamic decision making to invoke the suitable control actions iteratively for on-demand resource provisioning that delivers satisfactory user experience and controls cybersickness levels in a social VRLE.

## IV. PRIORITY-BASED QUEUING MODEL

In our 3QS-adaptation framework, the entire timeline of anomaly event processing can be divided into three parts each considered at *VRLE application plant*, *anomaly monitoring tool* and *decision module* as shown in Figure 2. This behavior of anomaly event data processing represents a queue, and thus we model our framework into a *M/M/1/K* finite queuing system to capture the pattern of VRLE application performance. This analytical model is based on an embedded Markov Chain, featured by states, events, transitions. The requests that enter into the queue are the anomaly events caused by 3QS issues, which are processed mainly on a priority basis i.e., in the order of events that have the ability to cause higher cybersickness levels. We focus especially on the response time in addressing the anomaly events inducing cybersickness. These anomaly events are processed on a single server hosted in our testbed setup detailed in Section V-A.

TABLE I: Performance metrics of our queuing model.

| No. of events in queue | $W_q$ (in sec) | $\bar{X}$ (in sec) | $R_s$ (in sec) | No. of processed severe anomalies |
|---|---|---|---|---|
| 10 | 2400.48 | 0.146 | 3300 | 4 |
| 20 | 5700 | 0.204 | 6303.98 | 5 |
| 30 | 8700 | 0.28 | 9304.15 | 7 |
| 40 | 11700 | 0.36 | 12304.32 | 11 |

The processing of an incoming request includes three stages: *stage 1* (collecting anomaly event data), *stage 2* (categorization into anomalies caused by 3QS issues), and *stage 3* (anomaly event data pushed into the decision module) as shown in Figure 3. After *stage 3*, the processed event record leaves the queue, where the anomaly data is sent to the decision module to determine the suitable action on the corresponding VRLE component. Each stage described in Figure 3, has a different average service rate, represented as $\mu_1$, $\mu_2$, and $\mu_3$. Thus, the overall response time of the system in processing one data record can be computed by solving the markov chain transition model. In this process, the execution of the three stages is mutually exclusive, which means that the second record will not be processed until the previous one is completed. We assume the processing times at each stage is exponentially distributed, and the data retrieval at stage 1 follow a Poisson arrival with an expected rate of $\lambda$.

The mean response time ($RT_q$) to process an anomaly event in the queue can be obtained by using the Little's formula [8].

The wait time of the queue $W_q$ is derived based on the number of events in the queue $L_q$ and arrival rate ($\lambda$)

$$W_q = \frac{L_q}{\lambda} \quad (2)$$

$\bar{X}$ is the sum of the mean service time for all three stages, and can be written as -

$$\bar{X} = \sum_{n=1}^{3} 1/\mu_n \quad (4)$$

We use the above analytical model in the performance evaluation experiments to determine the waiting delays that might occur in processing the anomaly events inducing cybersickness. To elucidate, a low cybersickness inducing anomaly trigger can be delayed, while a severe threat posing anomaly trigger can be urgently handled by allowing it to experience lower wait times in the triggers handling queue. To achieve such a handling, we use our priority queue model as a Binary-

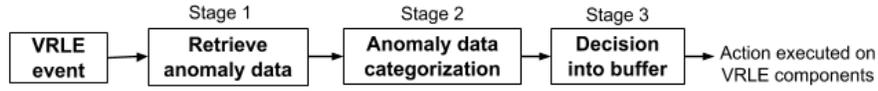

Fig. 3: Modeling stages of our proposed rule-based 3QS-adaptation framework as a queue.

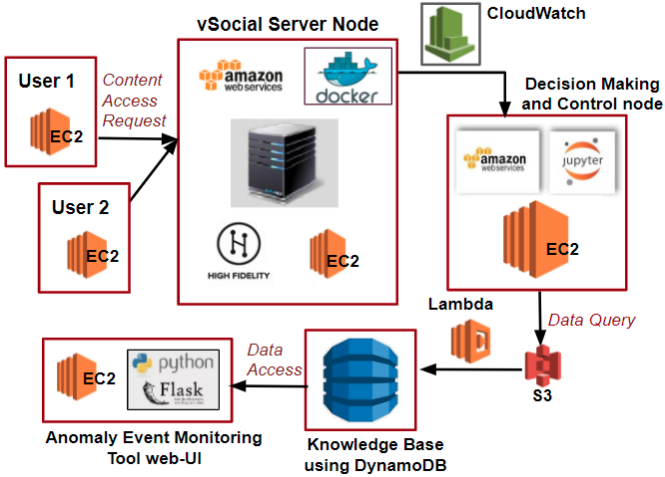

Fig. 4: Experimental testbed setup for 3QS-adaptation framework evaluation using anomaly event monitoring and rule-based decision making.

Heap [35] to perform reheapficiation of the events in the queue once an anomaly event is deleted from the queue.

Using the above formulation, Table I lists the calculation of the overall system response time ($R_s$) as Sum ($RT_q$, $R_{at}$), where $RT_q$ is the response time in queue and $R_{at}$ is the time taken for an adaptation to implement. In addition, we also enlist the number of processed severe anomaly events (i.e., with high cybersickness level) for a given number of anomaly events in the queue in Table I.

## V. Performance Evaluation

In this section, we first describe the experimental testbed setup for the evaluation of our rule-based 3QS-adaptation framework on the vSocial VRLE application case study [1]. Next, we discuss the results from experiments to perform a risk and cost aware analysis to finalize suitable adaptation choices. Lastly, based on our experimental results, we enlist suitable practices for handling the identified anomaly events in mitigating cybersickness within social VRLEs.

### A. Experimental Testbed Setup

We setup our experimental testbed in a public cloud i.e., Amazon Web Services (AWS) [36] as shown in Figure 4. In this testbed, we host the open-source vSocial application [1] on an Amazon Elastic Compute Cloud (EC2) instance [36] to render the VRLE content to the users. We also host a controller node on another EC2 instance to: (i) capture network data using Amazon CloudWatch [36], and (ii) monitor the network data using our anomaly monitoring tool alongside a decision module hosted on a separate Jupyter notebook instance [37]. In addition, we store the captured and processed network data in the controller node into a DynamoDB [36] service. This DynamoDB service serves as a knowledge base for future anomaly events. We also connect our knowledge base to Amazon S3 [36] service using the Amazon Lambda [36] service in order to provide seamless interaction between the decision module and the anomaly monitoring tool. Before illustrating our experimental scenarios, we first detail the tools used for anomaly data collection required for our framework.

As part of anomaly data collection, we simulate a QoS issue (packet drop), QoA issue (packet drop + network lag), Security issue (DoS, packet duplication + packet tampering) in our vSocial application setup. We calculate the packet rate by capturing the raw data associated to the timestamp of each packet for each of the simulated 3QS issues along with the baseline data (of benign behavior) of the vSocial application. To simulate a DoS attack on vSocial, we used Clumsy 0.2 [38], a windows based tool to control networking conditions such as lag, drop, throttle, or tamper of live packets. To see the impact on our VRLE application performance, we specifically drop a certain percentage of live packets. Using the Wireshark [39] tool, we capture packets being sent to-and-from our VRLE server in order to demonstrate possible data loss resulting from the packet capture. With the above specified tools and the experimental testbed setup, we collect the anomaly data relevant to 3QS issues in VRLE sessions.

### B. Anomaly Event Monitoring Tool

To identify traces of 3QS anomaly events in the collected network data during a VRLE session, we developed a web-based anomaly monitoring tool using the Flask micro framework with Python3 [40]. Our anomaly monitoring tool uses AWS CloudWatch alarms to create triggers based on a threshold condition for every 3QS anomaly. For instance, a QoS alarm is triggered if the threshold condition *if ([No. of packets out] < 7280 packets/second)* fails. Similarly, for a QoA alarm, we use a threshold condition if the (CPU Utilization %) > 8%. Next, the anomaly monitoring tool will pass the collected anomaly data to the decision module of our 3QS-adaptation framework as detailed in Section V-C. We store this detected anomaly data into a AWS S3 bucket [36], which is further interfaced with DynamoDB [36], the knowledge base.

### C. Adaptation Decision Making and Control

With the categorized anomaly data, the decision module will look up for the relevant decision unit as described in Section III-B. A sample list of potential adaptations for a specific decision unit are shown in the Table II. For example, a QoA issue arising due to {packet drop + lag} can be mitigated using the adaptations in Table II. Using the decision outcome, next the control module implements the adaptation based on the risk and cost aware analysis detailed in Section V-E.

TABLE II: Potential adaptation choices for different 3QS anomaly events.

| Anomaly Issue | Specific Category | Adaptation Name |
|---|---|---|
| QoA | High CPU Utilization | Upgrading Instance Type (A1) |
| | | Higher Resources (A2) |
| | | Modifying Instance Volume (A3) |
| QoS | Low Network Bandwidth | Enabling Enhanced Networking (A4) |
| | | Higher Network Bandwidth (A5) |
| Security | Denial of Service | Amazon Route 53 (A6) |
| | | AWS GuardDuty (A7) |
| Intrusion | Unauthorized Access | Blacklist IP via third-party app (A8) |

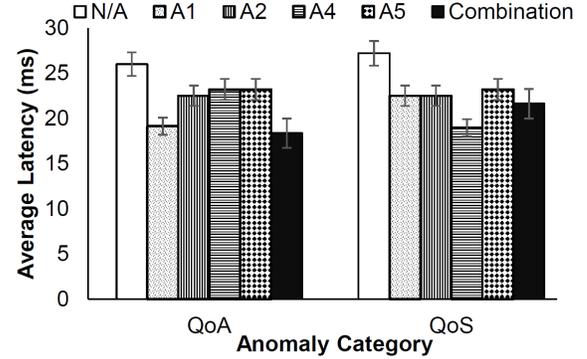

Fig. 5: Avg. Latency measured (in ms) for QoA anomaly, QoS anomaly scenario in different adaptation scenarios.

For instance, when an AWS alarm relevant to a QoA anomaly is triggered, the control module invokes an action for the suitable adaptation $A1$ keeping in mind the decision outcome, risk and cost factors. To elucidate, the control module upgrades the current instance (t2.micro) to c4.large based on the risk and cost aware analysis detailed in SectionV-E. After incorporating the adaptation A1, we observe there is a 4% decrease in CPU% as shown in Table III which is less than the threshold value. Similarly for a security issue, we utilize the adaptation Blacklist IP (A8) to block unauthorized access based on the threshold condition (number of login attempts $> 5$). Based on such implementations for anomaly events in VRLE, we show the results of our adaptations using the "performance metrics" {response time, Threshold measures}, cost incurred in Table III.

Once the decision related to an anomaly event is incorporated, its relevant information is updated into the knowledge base to train for future anomaly events. In our 3QS-adaptation framework, a knowledge base has been created using a DynamoDB service. To facilitate periodic updates from each of the modules in our framework into DynamoDB, we use the Amazon S3 service along with AWS Lambda functions. We use these both storage systems as our decision module that is hosted on a Jupyter notebook instance that takes only CSV data as input. The full capability of our knowledge base can be extended to other applications and can be utilized for employing additional adaptations in VRLE systems.

### D. Quantification of Cybersickness for 3QS Anomalies

In this section, we objectively measure the induced cybersickness level for a given set of anomaly events i.e., visualization delay due to network lag (QoA issue), packet loss (QoS issue), and DoS attack (security attack). Existing works [13], [14], [41] measure cybersickness based on physiological conditions (e.g., nausea, eyestrain). However, the works in [15] study that the quantifying effects of latency as the objective parameter to assess cybersickness. Based on these findings, we measure the latency as the primary objective metric of cybersickness for several 3QS anomaly events in VRLE. Note that each of these attack anomaly events are simulated in different network conditions as detailed in our prior work [6]. We also found that 23.5 ms is the baseline latency for a normal functioning VRLE session, beyond which a user experiences cybersickness.

The graphical results in Figure 5 detail the control of cybersickness (i.e., latency metric) level for the adaptations (i.e., upgrading instance (A1), scaling of higher Resources (A2), enhancing networking (A4), network bandwidth capacity (A5)) listed in Table II. We also consider a no-adaptation (NA) scenario to study the adverse impact on cybersickness if no action is taken to control the raised anomaly event. Moreover, in real-world applications such as vSocial, there is a possibility that one adaptation action might not be enough to mitigate the anomaly impact, and an adaptation should consider the possibility of a combination of performance and security issues inducing cybersickness [7].

To address such an case, we include a combination of adaptation scenarios to analyze the impact on cybersickness control for a given anomaly event. From the results in Figure 5, we observe that for a QoA anomaly, adaptations A1, A2 reduce the cybersickness by 26.43% and 13.46% respectively. In case of a QoS anomaly, the adaptation A4 reduces the cybersickness significantly by 30.28%. In addition, A1 and A2 reduce cybersickness by 17.28% making them the next suitable choice for a QoS anomaly as shown in Figure 5. We also note that the combination of best adaptations i.e., A1 and A4 reduces cybersickness by 29.39% for a QoA anomaly and 20.48% for a QoS anomaly event as shown in Figure 5. However the choice of combination can vary based on the considered list of potential candidates that can further impact the control of cybersickness levels in a VRLE session.

### E. Risk and Cost Aware Trade-off Analysis

*1) Risk calculation for the adaptations:* In this paper, we term risk as "failure risk" which is a likelihood value of an adaptation that can fail in controlling the cybersickness for a given anomaly event. We adopt the NIST SP800-30 [42] based risk assessment method [43] where we use $L(D)$– the likelihood of decision of a specific adaptation and $I$ represents the Impact of an adaptation in controlling the cybersickness level detailed in Figure V-D. We estimate the $L(D)$ based on the order of decision metrics. Using these both $L(D)$ and $I$, we calculate the failure risk as $R_f = 1 - f(L(D), I)$ where,

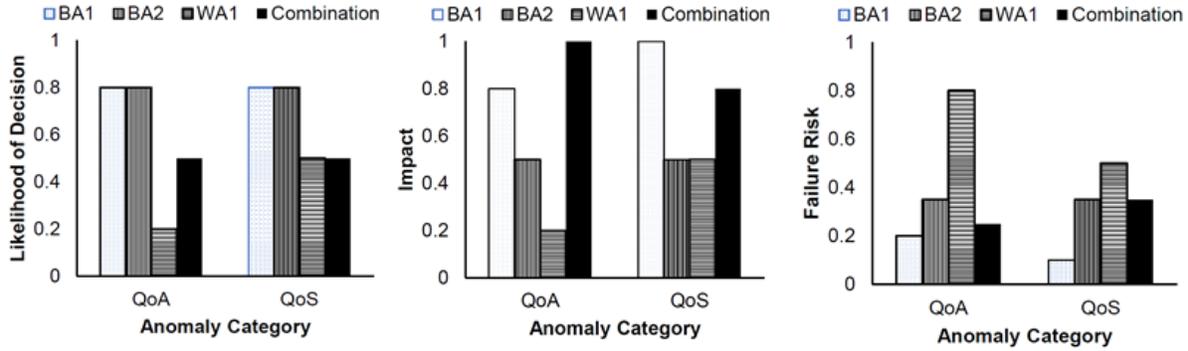

Fig. 6: Risk evaluation associated with the best (BA1, BA2), worst (WA1) and combination of adaptations in controlling cybersickness for the given QoA and QoS anomaly event.

$f(L(D), I)$ is the average function adopted from existing works [43]. We use a pre-defined semi-quantitative scale of 0-1 as guided by NIST for the impact/likelihood event assessments, with 1 indicating very high, and 0 indicating very low levels of impact. Using the latency measurement results in Figure 5, we consider the best/worst combination of adaptation choices for each anomaly event as illustrated in Figure 6.

*2) Cost aware application performance analysis:* The results in the previous section discussed the risk associated with some of the best and worst adaptations for a given set of anomaly categories in VRLEs. Each of these adaptations are implemented via the control action detailed in Section V-C. We measure the performance metrics and system response time due to these adaptations using CloudWatch as shown in Table III. With this, we highlight the functionality of our framework that takes dynamic decisions to control the cybersickness and maintain satisfactory application functionality. Based on our experimental results (i.e., cost-performance and risk evaluation), we enlist suitable rules (i.e., best practices) to adopt for future anomaly events.

*F. Recommendations Based on Key Findings*

Based on our initial experimental evaluation of our framework to control cybersickness level using the listed adaptations in Table II, we recommend rule-based practices as shown in Table IV. These practices are expressed in a semantic form i.e., we enlist Event-Condition-Action (ECA) rules with a typical form of $IF - THEN - (ELSE)$ [44] to adopt for future VRLE systems. From the results shown in Table IV in the context of a QoA anomaly event, we recommend adaptation A1 due to the low cost incurred in comparison to adaptation A2. In addition, from the risk results in Section V-E, we determine that adaptation A1 has a negative impact due to the high risk level in cybersickness control, when compared to adaptation A2. Moreover, for a QoS anomaly, we recommend adaptation A4 for the significant impact due to lower risk level on cybersickness over adaptation A5. Similarly, for an Unauthorized Access (UA), we recommend adaptation A8 over A7 due to the incurred cost and also the lack of control with the GuardDuty service in A7 as shown in Table IV. Moreover, the implementation of security recommendations in Table IV are aligned with the NIST SP800-160 security principles [16]. Specifically, mitigation strategies in: (i) A1 (e.g., addition of firewall rules, creation of security groups [18]) corresponds to the hardening principle, and (ii) A7 (e.g., addition of multiple components to increase impairment tolerance) corresponds to both hardening and diversity principles. In addition, our recommendations can range from ideas of checking for malware and updating security groups to extreme actions such as terminating the application instance altogether. With our proposed approach, to avoid the false alarms for performing an adaption, we first determine the cybersickness levels which is based on the objective parameters. In addition, in the event of low accuracy levels, our solution focuses on maintaining the functionality of the VRLE application. Using such rule-based adaptations, we showcase the benefit of our proposed framework that controls the cybersickness level induced by the 3QS related anomalies.

## VI. CONCLUSION

Social VRLEs provide immersive experience to the users via remote learning capabilities. The current lack in the state-of-the-art to address potential performance/security issues can disrupt users' experience and potentially induce cybersickness. Failure to address such issues can endanger the user safety by causing physical harm e.g., by obstructing the view of the users, forcing users to run into walls. In this paper, we proposed a novel rule-based 3QS-adaptation framework that focused on the control of cybersickness induced by performance issues (e.g., visualization delay, packet drop, time lag) and security issues (e.g., unauthorized access, DoS attack). We quantified the cybersickness metric objectively using a latency metric for a simulated anomaly event scenario. We utilized a priority-based queuing model that handles anomaly events in the order of highest cybersickness inducing levels. To determine the suitable adaptation for handling a given anomaly event type, our approach involves performing risk and cost aware analysis for each decision outcome. Once a suitable adaptation is incorporated for a given anomaly event type, cybersickness measurements are updated and used as feedback to determine the impact on the anomaly event.

TABLE III: Cost-aware application performance analysis of adaptations chosen for 3QS anomaly events.

| Anomaly Event | Adaptation name | Cost (in $/hr) | Threshold Metric | $R_{at}$ (in seconds) |
|---|---|---|---|---|
| QoA | A1 | 0.23 | CPU utilization rate | 0.54 |
| | A2 | 2.4 | is decreased to 4% | 300 |
| QoS | A4 | 0.10 | Packet rate at 7280 | 1 |
| | A5 | 0.10 | packets/second | 300 |
| DoS | A7 | 0.33 | Packet data measure | 0.51 |
| Unauthorized access | A8 | 0.02 | Number of login attempts <5 | Varies based on number of users |

TABLE IV: Recommendations based on risk level ($R_l$), Cost level ($C_l$), and control on cybersickness ($\Delta CS$).

| IF | | THEN | | | | ELSE | | | |
|---|---|---|---|---|---|---|---|---|---|
| Anomaly | Scenario in VRLE session | $A_i$ | $R_l$ | $C_l$ | $\Delta CS\%$ | $A_i$ | $R_l$ | $C_l$ | $\Delta CS\%$ |
| QoA | Increasing number of users; To improve application run time | A1 | L | L | 26.43% | A2 | M | M | 13.46% |
| QoS | Lower latency in VRLE content | A4 | L | L | 30.28% | A1+A4 | L | M | 20.48% |
| UA | Only valid users in VRLE session | A8 | L | L | 20.7% | A7 | M | H | - |
| DoS | Avoid loss of content availability | A1+A6 | M | M | 36.1% | A1+A7 | M | H | - |

Our validation results show that the real-time adaptations suggested by our rule-based framework: (i) reduce the cybersickness level by 26.43% for a QoA anomaly and the same for a QoS anomaly event by 30.28%, and (ii) maintains the application functionality within the threshold limit (beyond which an application is non-functional) along with low system response times. Based on these key findings, we enlisted suitable practices for prevention of 3QS issues based on NIST SP800-160 guidelines.

As part of future work, we plan to study issues around identifying zero-day anomalies and performing more extensive adaptations in real-time. These adaptations could feature refined rules for dynamic decision making, as and when a VRLE is scaled to handle more number of session users.